\documentclass[a4paper,12pt]{article}
\usepackage{amsmath,amsfonts,amssymb}
\usepackage[latin1]{inputenc}
\usepackage[english]{babel}
\usepackage{graphicx}
\usepackage{rotating}
\usepackage{latexsym}
\usepackage{cite}
\usepackage{subfigure}

\newcommand{\be}{\begin{eqnarray}}
\newcommand{\en}{\end{eqnarray}}

\begin{document}

\thispagestyle{empty}

\vspace*{20mm}
\begin{center}
\textbf{ \Large{Symmetry breaking induced by 't Hooft non-abelian flux} }

\vspace{1.0cm}
M. Salvatori 

\vspace{1.0cm}
{\textit{Departamento de F\'{\i}sica Te\'{o}rica  and Instituto
 de F\'{\i}sica Te\'{o}rica , Universidad Aut\'{o}noma de Madrid,
 Cantoblanco, E-28049 Madrid, Spain.}}\\[0pt]
\end{center}

\vspace{2cm}
\begin{center}
{\bf Abstract}
\end{center}
We analyze a  $SU(N)$ gauge theory on  $\mathcal{M}_4 \times T^2$.
We find and catalogue all possible degenerate zero-energy stable configurations in the case of trivial or non-trivial 't Hooft non abelian flux. We describe the residual symmetries of each vacua and the $4$-dimensional effective spectrum in terms of continous and discrete parameters, respectively.

\clearpage


\section{Introduction}

In the Standard Model, the explanation of the electroweak symmetry breaking and the correlated Hierarchy problem still remains a non completely understood  issue. 
Theories with extra space-like dimensions compactified on non-simply connected manifolds represent a possible scenario where trying to solve this problem. Such context, indeed, offers a new  possibility of symmetry breaking: the Scherk-Schwarz (SS) mechanism \cite{ss}.  The non local nature of this symmetry breaking protects the theory from  ultraviolet divergences  and makes it a promising candidate mechanism to break  the electroweak symmetry. In this paper we analyze the $SU(N)$ stable vacua and their symmetries in the case of two extra space-like dimensions  compactified \textit{\'a la} SS on a 2-dimensional torus, $T^2$. The case of trivial non-abelian t'Hooft flux is well-known in the literature \cite{Hosotani}, while the phenomenology of the non-trivial non-abelian t'Hooft flux has not been explored yet.

\section{$SU(N)$ on $\mathcal{M}_{\mathrm{4}} \times T^{\mathrm{2}}$: stable vacua and symmetries}

Consider a $SU(N)$ gauge theory on  a space time of the type $\mathcal{M}_{\mathrm{4}} \times T^{\mathrm{2}}$. We will denote by $x$ the  coordinates of the $4$-dimensional Minkowski space $\mathcal{M}_4$ and by $y$ the extra-space like dimensions compactified on a $2$-dimensional torus. 

A gauge field living on $\mathcal{M}_{\mathrm{4}} \times T^{2}$  has to be periodic up to a gauge transformation under the fundamental shifts $\mathcal{T}_a : y \rightarrow y + l_a$ with $a=1,2$,  that define the torus:
\begin{equation}
\begin{array}{ccc}
\mathbf{A}_M (x,y + l_a) &=& \Omega_a(y) \mathbf{A}_M (x,y) \Omega_a^\dagger(y) + \frac{i}{g} \Omega_a(y) \partial_M \Omega_a^\dagger(y) 
\label{boun_cond_A} \\
\mathbf{F}_{MN}(x,y+l_a) &=& \Omega_a(y) \mathbf{F}_{MN}(x,y) \Omega^\dagger(y) \,,
\end{array}
\end{equation}
where $M,N=0,1,...,5$, $a=1,2$ and $l_a$ is the length of the direction $a$. The eq.\eqref{boun_cond_A} is known as coordinate dependent Scherk-Schwarz compactification. The transition functions $\Omega_a(y)$ are the embedding of the fundamental shifts in the gauge space and 
in order to preserve $4$-dimensional Poincar\'e invariance, they can only depend on the extra dimensions $y$.
Under a gauge transformation $S \in SU(N)$, the $\Omega_a(y)$ transform as $
\Omega_a' (y) = S(y+l_a) \, \Omega_a(y) \,S^\dagger(y)$. In addition, the transition functions  are constrained by the following consistency condition coming from the geometry:
\begin{equation}
\Omega_1(y+l_2)\,\Omega_2(y)\,=\,e^{2 \pi i \frac{m}{N}}\,\Omega_2(y+l_1)\, \Omega_1(y) \,.
\label{cons_cond}
\end{equation}
The factor  exp$[2 \pi i m/N]$ is the embedding  of the identity in the gauge space\footnote{A non-trivial value of $m$ is possible only in absence of field representations sensitive to the center of the group.} and the gauge invariant quantity $m$ is a topological quantity called \textit{non-abelian 't Hooft flux} \cite{'tHooft:1979uj}.

For a generic Lie gauge group $\mathcal{G}$ on $T^{2n}$, all sets of transition functions $\{\Omega_a\}$ satisfying eq.\eqref{cons_cond} are \textit{gauge equivalent}  if and only if $\mathcal{G}$ is $2n-1$-connected, that is, if $\Pi_i(\mathcal{G})=0$ $\forall i=1,..,2n-1$ \cite{salva}. In particular, when $n=1$ and $\mathcal{G}=SU(N)$, \textit{i.e.} a simple connected gauge group on a $2$-dimensional torus, all  solutions of  eq.\eqref{cons_cond} are \textit{gauge equivalent} \cite{salva,amb}. 

For any choice of $\Omega_a$ satisfying the constraint in eq.~(\ref{cons_cond}), it is always possible, therefore, to find a $SU(N)$ gauge transformation $U(y)$ satisfying 
\begin{equation}
U(y + l_a) \,\,= \,\,\Omega_a(y) \,U(y)\,V_a^\dagger \hspace*{1cm} \mathrm{with} \,\,\,a=1,2\,,
\label{periodicity_U}
\end{equation}
where $V_a$ are constant solutions of the consistency condition in eq.(\ref{cons_cond}). 

Two pairs $V_1, V_2$ and $V_1', V_2'$ are called \textit{non-equivalent} if they are  not connected by a $SU(N)$ gauge transformation. Eq.~\eqref{periodicity_U} allow to show  \cite{salva,ABGRS} that, given the transition functions $\Omega_1(y), \Omega_2(y)$, for each \textit{non-equivalent} pair of $V_1, V_2$ there exists a different gauge transformation $U(y)$ giving rise to a different stable zero-energy configuration of the type 
\begin{equation}
\langle \mathbf{A}_a \rangle \,\,= \,\,\frac{i}{g}\,U(y) \,\partial_a\,U^\dagger(y) \,,
\end{equation}  
compatible with the periodicity conditions in eq.(\ref{boun_cond_A}).
To classify the possible stable vacuum configurations for a $SU(N)$ gauge theory on $\mathcal{M}_{\mathrm{4}} \times T^{\mathrm{2}}$ with a fixed value of the t' Hooft non abelian flux $m$, it  means, thus, to catalogue all possible \textit{non-equivalent} pairs of $V_a$. This result is even more transparent in the \textit{background symmetric gauge} \cite{salva}, that is the gauge in which $\langle\mathbf{A}_a^{sym}\rangle = 0$ and $\Omega_a^{sym}= V_a$. 

\begin{table}
\begin{tabular}{lcl}
\hline 
 & \vline &\\
\multicolumn{1}{c}{$\mathbf{m\neq0}$} & \vline &  \multicolumn{1}{c}{\textbf{m=0}}   \\
& \vline &\\
\hline
& \vline & \\
$V_a\,=\,P^{\alpha_a}\,Q^{\beta_a}$  & \vline & $V_a = e^{2 \pi i \alpha_a^j H_j}$   \\
 & \vline & \\
$\begin{array}{l}
P_{kj}=e^{- 2 \pi i \frac{(k-1)}{N}} \,\,e^{i \pi \frac{N-1}{N}}\,\, \delta_{kj} \\
Q_{kj}= e^{i \pi \frac{N-1}{N}} \,\, \delta_{k,j-1}
\end{array}$ & \vline & 
$\begin{array}{l}
H_j \in \mathrm{Cartan} \hspace*{1em} \left[H_i, H_j\right]=0 \\
i,j=1,...,N-1
\end{array}$ \\
 & \vline & \\
$\begin{array}{l}
\alpha_1\,\beta_2-\alpha_2\,\beta_1=m \\
|\alpha_a|, |\beta_a| = 0,1,..,N-1 
\end{array}$ & \vline & $\alpha_a^j \,\in [0,1[$ \\
\hline
 & \vline & \\
$SU(N) \rightarrow SU(k_2)^{\frac{k_1}{k_2}} \times U(1)^{\frac{k_1}{k_2}-1}$  & \vline & $\forall \alpha_a^j \neq 0$ \hspace*{2em}
$SU(N) \rightarrow U(1)^{N-1}$ \\
 & \vline & \\
\hline
\end{tabular}
\caption{Constant solutions of eq.\eqref{cons_cond} and their symmetries. $k_1= g.c.d.(m,N)$, $k_2= g.c.d.(\alpha_1,\alpha_2,\beta_1,\beta_2,N)$.}
\label{tab:a}
\end{table}
The constant solutions  of the consistency condition in eq.\eqref{cons_cond} and the residual symmetries for non-trivial and trivial 't Hooft non abelian flux $m$ are summarized in table \ref{tab:a}. 

For $m=0$, $V_a$ commute and they can only depend on the $N-1$ generators belonging to the Cartan subalgebra  of $SU(N)$. The constant transition functions $V_a$ are completely characterized by  $2(N-1)$ parameters ($\alpha_a^j$) that take values continuously between 0 and 1. Such parameters are non-integrable phases, which arise only in a topologically non-trivial space and cannot  be gauged-away. Their values are determined dynamically only at quantum level.
Since  $V_a$ commute, the symmetry breaking is rank-preserving, resulting in the following $4$-dimensional mass spectrum:
\begin{equation}
\begin{array}{lcr}
m_{n_1,n_2}^j = 4 \pi^2 \left[ \frac{n_1^2}{l_1^2} +  \frac{n_2^2}{l_2^2} \right] & \mathrm{if} & \mathrm{\mathbf{A}_\mu^j} \in \mathrm{Cartan} \\
m_{n_1,n_2}^r = 4 \pi^2 \,\sum_{a=1}^2\,\left[ \left(n_a + \sum_{j=1}^{N-1} \,q^j_r \alpha_a^j\,\right)^2 \frac{1}{l_a^2} \right] & \mathrm{if} & \mathrm{\mathbf{A}_\mu^r} \in\!\!\!\!\!\!\slash \,\,\mathrm{Cartan}\,. 
\end{array}
\label{kk_spectrum}
\end{equation}
$n_1, n_2$ are integers and $q_r^j$ are roots of $SU(N)$. When all $\alpha_a^j \neq 0$, the only $4$-dimensional massless gauge bosons are those associated to the Cartan subalgebra, giving rise to the maximal rank-preserving symmetry breaking $SU(N)\rightarrow U(1)^{N-1}$.

In the case $m\neq0$,  $V_a$ do not commute. There exists a finite number of constant solutions of eq. \eqref{cons_cond} described by 4 discrete parameters ($\alpha_a, \beta_a$) that take integer values in the range $\left[-N+1, N-1\right]$. These parameters are subject to the following constraint coming from  eq.\eqref{cons_cond}: $\alpha_1 \beta_2 - \alpha_2 \beta_1 = m$. Notice that $\alpha_a, \,\beta_a$ cannot be simultaneously zero. 

Since the transition functions do not commute, the residual symmetry group has  rank lower than $SU(N)$. In particular, in terms of the two parameters\footnote{g.c.d.= great common divisor.} $k_1= g.c.d.(m,N)$ and $k_2=g.c.d.(\alpha_1, \alpha_2, \beta_1, \beta_2, N)$, the symmetry breaking reads $SU(N) \rightarrow SU(k_2)^{\frac{k_1}{k_2}} \times U(1)^{\frac{k_1}{k_2}-1}$ \cite{salva,ABGRS}. 
Notice that for a given $m$ and $N$ (and consequently $k_1$), it is possible to have different degenerate vacua characterized by different sets of discrete parameters $\alpha_1,\,\alpha_2,\,\beta_1,\,\beta_2$. They correspond to  different values of $k_2$ and therefore different residual symmetries. Only quantum effects remove such degeneration and determine the true vacuum of the theory. The $4$-dimensional gauge mass spectrum reads \cite{salva} 
\begin{equation}
\label{m_spectrum_1}
m^2_{n_1,n_2} = 4 \pi^2 \,\sum_{a=1}^2\,\, \left(n_a + \frac{\alpha_a\,\Delta\,\,+\,\,\beta_a\,k_\Delta}{N} \right)^2 \frac{1}{l_a^2} \,,    
\end{equation}
where $n_1,n_2$ are integer numbers, $\Delta=0,1,...,N-1$, and $k_{\Delta=0} =1,...,N-1$ and $k_{\Delta\neq0}=0,1,...,N-1$. The spectrum in eq.\eqref{m_spectrum_1} manifests the symmetry breaking of table \ref{tab:a}: the only zero modes are associated to the $k_1\,k_2-1$  $4$-dimensional gauge bosons for which $\frac{\alpha_a\,\Delta\,\,+\,\,\beta_a\,k_\Delta}{N}$ ($\forall a=1,2$) are integer numbers.

Notice that the spectra for the case $m=0$ and $m\neq0$ in eq.\eqref{kk_spectrum} and  eq.\eqref{m_spectrum_1} respectively, show a similar structure with the only difference that the symmetry breaking masses are expressed in terms of  continous ($m=0$) and discrete ($m\neq0$) parameters. While in the $m=0$ case,  the scale of the lightest non-zero masses  $2 \pi \alpha_a/l_a$ with $a=1,2$ is arbitrary and it is fixed only at the quantum level,  for the $m\neq0$ case the non trivial constraint in eq.\eqref{cons_cond} determines the new  scales  $\frac{2 \pi}{l_a} \,\frac{1}{N}$ already at the classical level.

Finally, it is worth to  underline the different nature of  the symmetry breaking for  the two cases of trivial ($m=0$) and non trivial ($m\neq0$)'t Hooft non abelian flux. In the case $m=0$, indeed, the gauge symmetry breaking is exactly like the Hosotani mechanism \cite{Hosotani}: it is always possible to choose an appropriate background gauge, compatible with the consistency conditions, in which the transition functions are trivial ($V_1=V_2=\mathbf{1}$) and the extra space-like components of the six-dimensional gauge fields $\mathbf{A}_a$ acquire a vacuum expectation value (VEV): $\langle\mathbf{A}_a\rangle=B_a$. In this case, the symmetry breaking can be seen as spontaneous in the following sense:
\begin{enumerate}
\item For each $4$-dimensional massive gauge field $\mathbf{A}_{\mu}$, there exists a linear combination of the $\mathbf{A}_a$ that play the role of a $4$-dimensional scalar pseudo-goldstone boson eaten, by the $4$-dimensional gauge bosons to became a longitudinal degree of freedom. 
\item The VEV of $\mathbf{A}_a$ works as the order parameter of the symmetry breaking mechanism. In particular, it is possible  to deform $\langle\mathbf{A}_a\rangle$ to zero compatibly with the consistency conditions  so as to restore all the initial symmetries. 
\end{enumerate}
In the case $m \neq 0$, we cannot interpret the symmetry breaking mechanism as a spontaneous symmetry breaking mechanism. The consistency conditions, indeed, forbid to have trivial transition functions and then \textit{the symmetry breaking can not be related \textbf{only} to the VEV of} $\mathbf{A}_a$. Although for each massive $4$-dimensional gauge boson $\mathbf{A}_{\mu}$ there exists a $4$-dimensional pseudo-goldstone boson, it is not possible to determine an order parameter that can be  deformed, compatibly with the consistency conditions, in such a way to restore all the initial symmetries.

\section*{Acknowledgements}
I acknowledge for very interesting discussion to J. Bellorin, A. Broncano, A. Ruzzo, M.B. Gavela and J. Alfaro. I am special indebted for the discussion and for reading the manuscript to S. Rigolin.  I also acknowledge MECD for financial support through FPU fellowship AP2003-1540.

\bibliographystyle{aipproc}






\end{document}